# Algorithm and Architecture for Hybrid Decoding of Polar Codes

Bo Yuan and Keshab K. Parhi, *Fellow*, *IEEE*
Department of Electrical and Computer Engineering, University of Minnesota Twin Cities

***Abstract*—Polar codes are the first provable capacity-achieving forward error correction (FEC) codes. In general polar codes can be decoded via either successive cancellation (SC) or belief propagation (BP) decoding algorithm. However, to date practical applications of polar codes have been hindered by the long decoding latency and limited error-correcting performance problems. In this paper, based on our recent proposed early stopping criteria for the BP algorithm, we propose a hybrid BP-SC decoding scheme to improve the decoding performance of polar codes with relatively short latency. Simulation results show that, for (1024, 512) polar codes the proposed approach leads to at least 0.2dB gain over the BP algorithm with the same maximum number of iterations for the entire SNR region, and also achieves 0.2dB decoding gain over the BP algorithm with the same worst-case latency in the high SNR region. Besides, compared to the SC algorithm, the proposed scheme leads to 0.2dB gain in the medium SNR region with much less average decoding latency. In addition, we also propose the low-complexity unified hardware architecture for the hybrid decoding scheme, which is able to implement SC and BP algorithms using same hardware.**

## I. INTRODUCTION

Polar codes have received much attention from information theorists because of their capacity-achieving property [1]. Successive cancellation (SC) [1] and belief propagation (BP) [2] algorithms are the two main decoding approaches for polar codes. However, to date the practical applications of polar codes are impeded by their long decoding latency and limited error-correcting performance problems. Prior investigations have proposed solutions to overcome these problems at both algorithm and VLSI implementation levels. In [3-5], SC list (SCL) algorithm and its variants were proposed to improve the performance of polar codes. Meanwhile, [6-10] presented several approaches to reduce the decoding latency of SC and SCL decoders. Besides, improvements in decoding performance of BP algorithm were reported in [11-14]. However, even with the use of the above efforts, polar codes have not still met the requirements of very short latency and improved decoding performance with short critical path delay.

Based on our recent progress on early stopping criteria for a BP decoder, in this paper we propose a new BP-SC hybrid decoding scheme. Different from an $L$-size SCL decoder, the proposed hybrid decoder only consists of one *preprocessing* BP decoder and one SC decoder. With the help of preprocessing BP decoder, the average decoding latency of entire decoding is significantly reduced. On the other hand, the *denoised* channel information output from the preprocessing BP decoder can also help to improve the error-correcting performance of the SC component decoder. As a result, the entire hybrid decoder achieves a much shorter decoding latency with improved decoding performance. For (1024, 512) polar codes, simulation results show that the proposed approach leads to at least 0.2dB gain compared to a conventional BP decoder with the same maximum number of iterations for the entire SNR region. In addition, the proposed decoder also achieves 0.2dB gain over the BP decoder with the same worst-case latency in the high SNR region. Besides, compared to a traditional SC decoder, the proposed approach leads to a 0.2dB gain in the medium SNR region with much less average decoding latency. Moreover, we also develop the low-complexity unified hardware architecture of the hybrid decoder, which can implement SC and BP algorithms on the same hardware.

The rest of this paper is organized as follows. Section II gives a brief review of polar codes as well as SC and BP algorithms. The proposed hybrid decoding scheme is presented in Section III. Analysis and comparison on latency and decoding performance are also discussed in this section. Section IV presents the unified hardware architecture for the hybrid decoding scheme. Section V draws the conclusions.

## II. REVIEW OF POLAR CODES

### A. *Encoding Process of Polar Codes*

Polar codes were proposed based on the polarization of reliability of decoded bits [1]. In general, the positions of reliable and unreliable decoded bits are referred as free and frozen positions, respectively. Fig. 1 shows an example polar encoding procedure. First the length-$k$ source data is extended to $n$-length vector $\boldsymbol{u}$ by assigning frozen positions as "0". Then $\boldsymbol{x}$=$\boldsymbol{uG}$ is transmitted over the channel, where $\boldsymbol{G}$ is the $n \times n$ generator matrix. Finally $\boldsymbol{x}$ is transmitted as the transmitted codeword. For the details of polar encoding process, the reader is referred to [1].

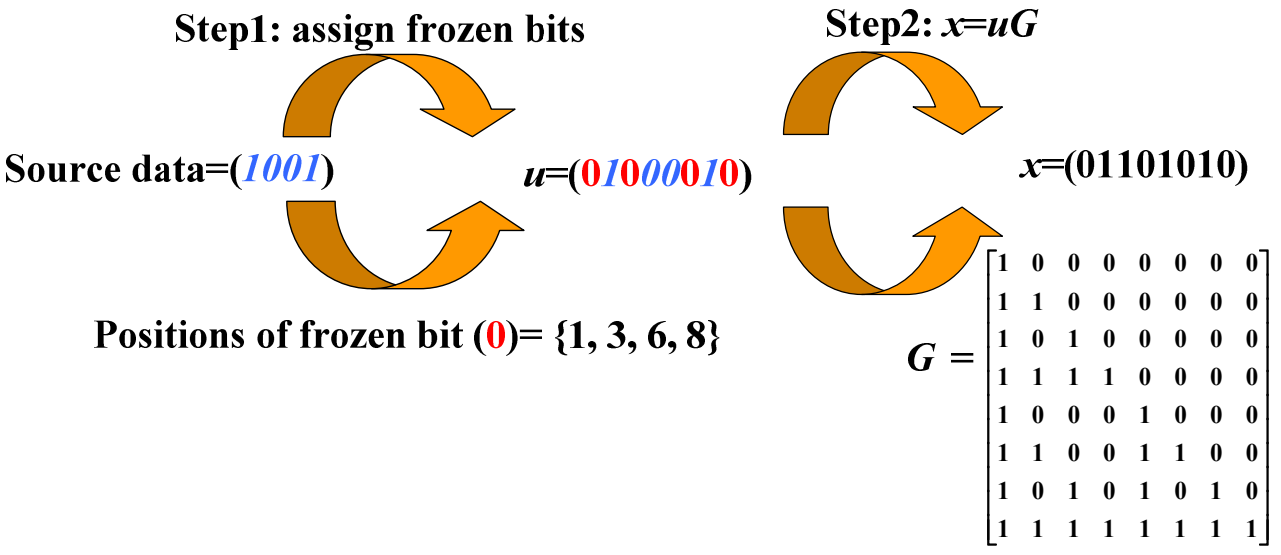

Fig. 1. Example polar encoding with $n$=8 and $k$=4.

## B. Decoding Process of Polar Codes

At the receiver end, due to the channel noise, ***x*** is corrupted to the received codeword ***r***. An SC or BP decoder is used to recover ***u*** from ***r***. Fig. 2 shows the example SC decoding procedure for *n*=8 polar code. Here the entire SC decoder consists of two types of processing nodes, namely **f** and **g** nodes. Each node is associated with a number that indicates the index of the clock cycle when the node is activated. In addition, a hard-decision **h** unit is used to determine the decoded bit. From Fig. 2 it can be seen that the decoding scheme of SC decoder is serial, which causes long latency. For details of SC decoding, the reader is referred to [9].

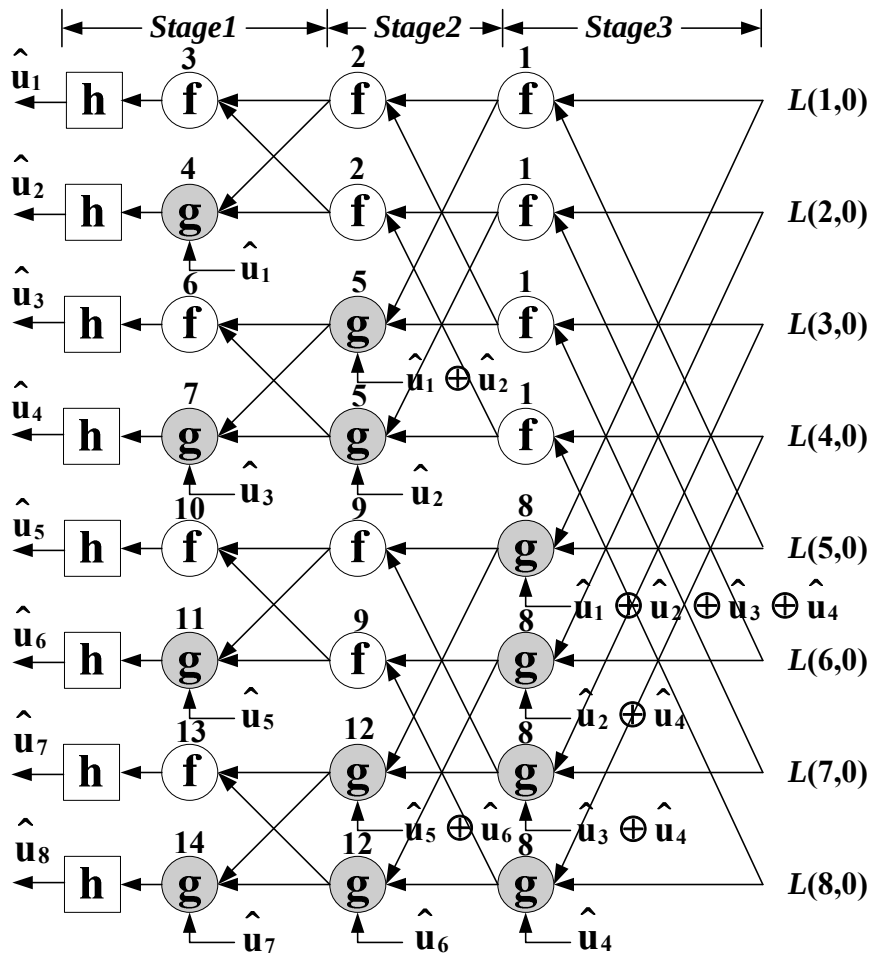

Fig. 2. Example polar SC decoding with *n*=8, cited from [9].

Fig. 3 shows the example BP decoding procedure for polar codes with *n*=8. Similar to an SC decoder, the BP decoder consists of $\log_2 n$ stages as well, where each stage contains the check and equity nodes. Different from the serial-output SC decoder, the BP decoding is iterative and inherently parallel. For details of BP decoding, the reader is referred to [12].

## C. Early Stopping Criteria for BP Decoder

For many applications, the iterative decoder can find the valid codeword before reaching the pre-set maximum number of iterations referred as *max iter*. In those cases, the early stopping criteria are very important because they help terminate the iterative procedure early to save decoding time and energy. In our recent work [15], various early stopping criteria for polar BP decoder were proposed to reduce the average number of iterations. For example, for (1024, 512) polar codes transmitted over AWGN channel with BPSK modulation, the average required number of iterations is only 26.1 at SNR=3dB, which is much less than the conventional pre-set *max iter*=60. As a result, the entire decoding latency and energy are reduced significantly.

Considering the great benefit offered by the early stopping criteria, this paper integrates this technique into the proposed hybrid decoding scheme to develop low-latency improved performance polar decoder, which is presented in Section III.

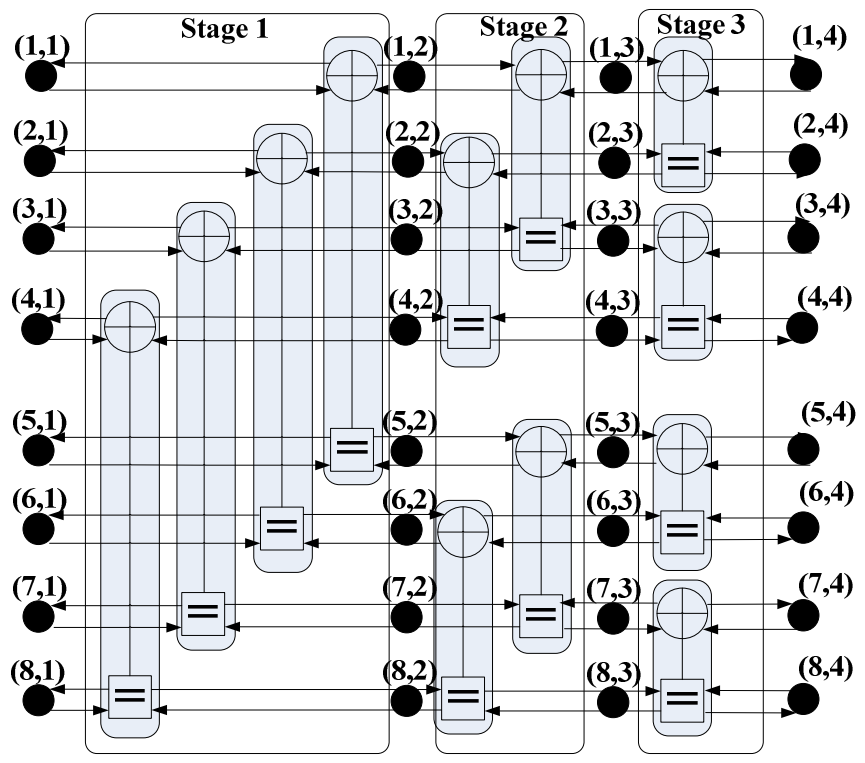

Fig. 3. Example polar BP decoding with *n*=8, cited from [12].

# III. The Proposed Hybrid Decoding Approach

## A. BP-SC Hybrid Decoding Scheme

Fig. 4 shows the overall architecture of the proposed hybrid polar code decoding scheme. It can be seen that this hybrid decoder consists of one front-end preprocessing BP decoder and one back-end SC decoder. Here the back-end SC decoder is a reformulated *8-bit output decoder* that was proposed in [10]. In the first phase of hybrid decoding, the input channel log-likelihood ratios (LLRs) are sent to the iterative BP decoder. At the end of each iteration, the early stopping criteria that were devised in [15] detect whether the valid decoded codeword has been found or not. If the valid codeword is found, then the iteration is terminated and the BP decoder outputs the valid codeword. In that case the SC decoder is not activated at all during the entire procedure. If the valid codeword is not found at this round of iteration, then the entire iterative procedure continues. If the front-end BP decoder is unable to decode the valid codewords even reaching *max iter*, then it outputs *denoised* channel LLRs to the back-end SC decoder to perform extra SC decoding. In that case the final decoded codeword is output from the SC decoder instead of the BP decoder.

Notice that the key part of the proposed hybrid decoding scheme is the use of denoised channel LLRs output from BP decoder. This approach is based on the special property of BP decoder that it can generate two types of LLR information at its two ends. After some rounds of iteration, the LLRs for estimated ***u*** are available at the left end of BP decoder (see Fig. 3); meanwhile, the LLRs for estimate ***x*** are also available at the right end of BP decoder. As discussed in [15], these LLRs for estimated ***x*** are less noisy than the original channel LLRs for ***x***. This is because part of the noise that is contained in the channel LLRs has been removed during the iterative BP procedure. Therefore, these denoised LLRs, instead of the original channel LLRs, are sent to the back-end SC decoder to decode the codeword.

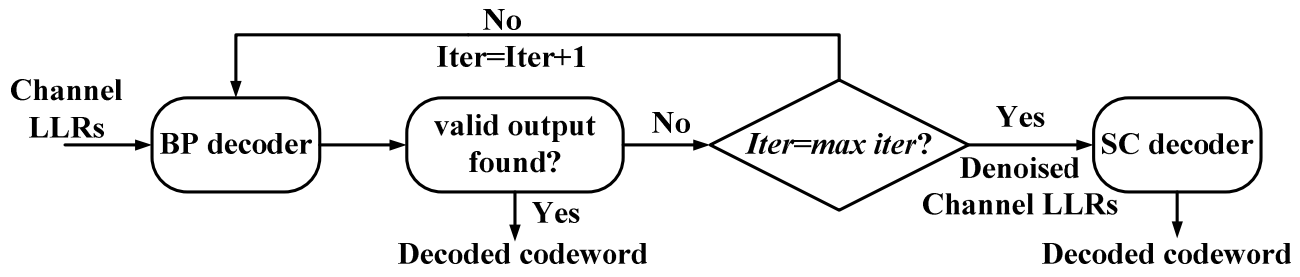

Fig. 4. Hybrid polar decoding scheme.

## B. Simulation Results and Latency Analysis

Fig. 5 shows the decoding performance of the proposed hybrid decoder for (1024, 512) polar codes. Here the *max iter* for the front-end BP decoder is set to 60. In addition, the performance of single BP and SC decoders are also provided in the figure for fair comparison.

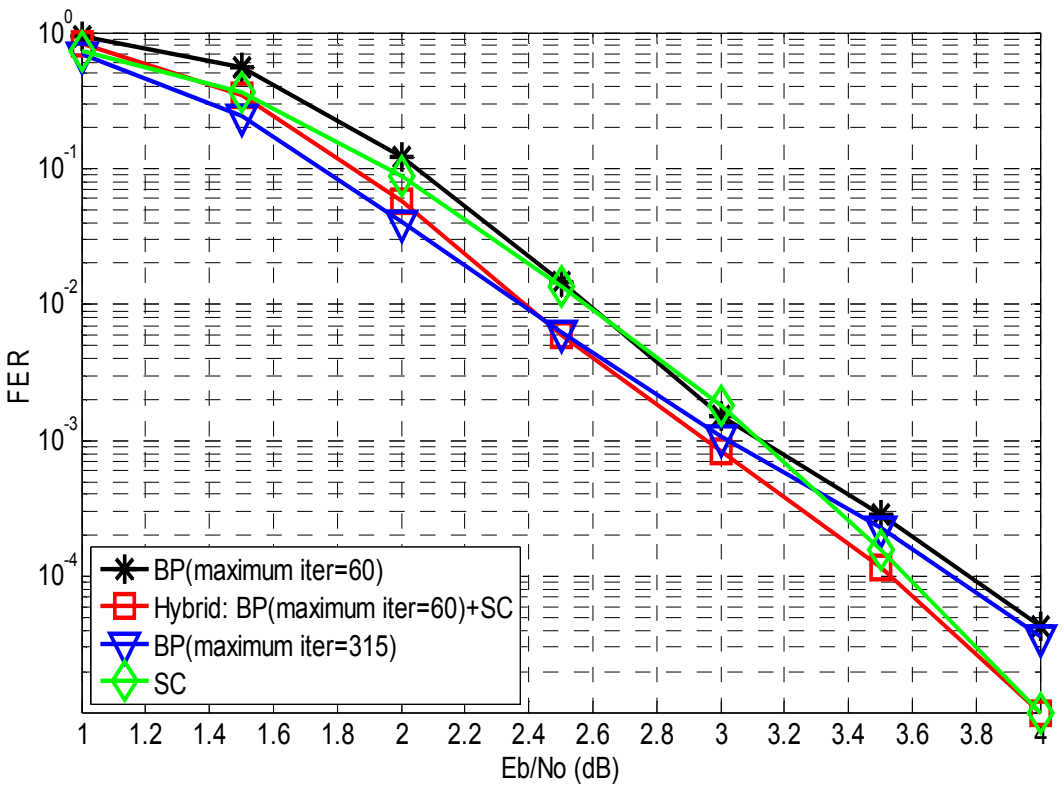

Fig. 5. Performance of hybrid decoding for (1024, 512) polar codes.

As shown in Fig. 5, the hybrid decoder outperforms BP decoder with the same *max iter* by at least 0.2dB for the entire SNR region. This is because for some codewords that cannot be decoded via a single BP decoder, the back-end SC decoder in the hybrid decoder is able to decode them with the use of denoised channel LLRs. In addition, even compared to the single BP decoder with very large *max iter* (for example 315 in Fig. 4), the hybrid decoder still outperforms it with 0.2dB in high SNR region. This indicates that for the cases that increasing *max iter* cannot bring sufficient decoding gains, the proposed hybrid decoder can still lead to significant improvement on error-correcting performance. Besides, Fig. 5 shows that the hybrid decoder can also achieve 0.2dB gain over an SC decoder in medium SNR region.

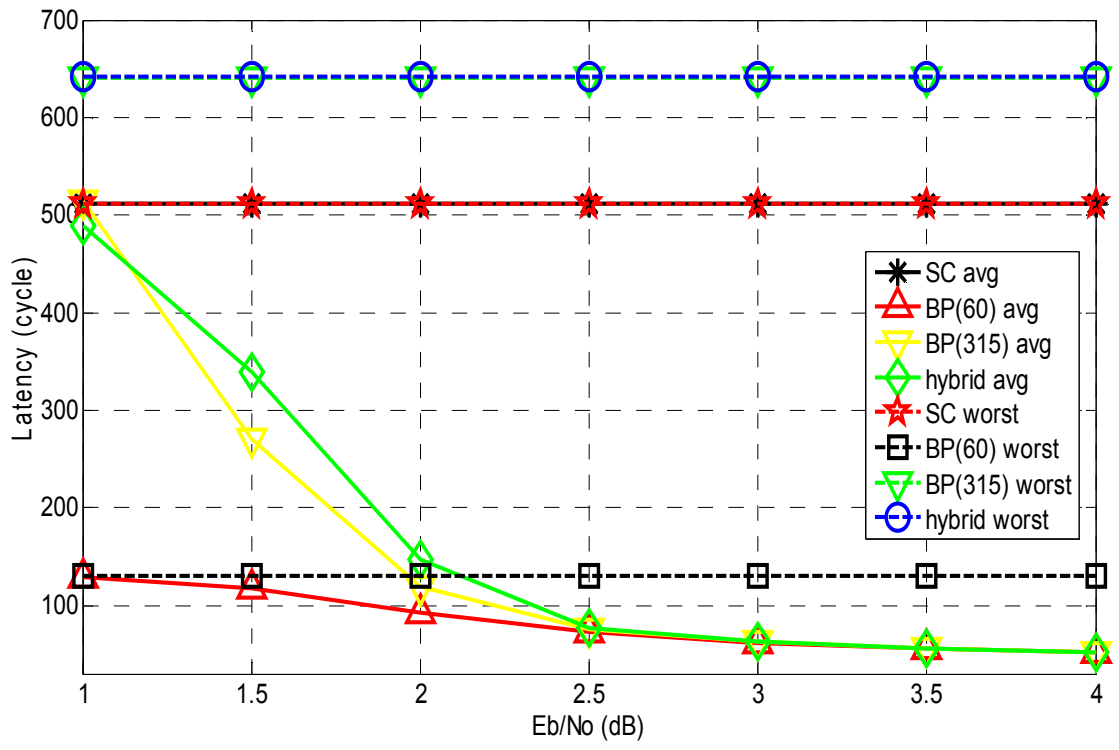

Fig. 6. Latency of hybrid decoding for (1024, 512) polar codes.

Fig. 6 shows the decoding latency of different decoders in terms of clock cycles. It can be seen that compared to the single BP decoder with *max iter*=60, the proposed hybrid decoder has larger worst-case latency; however, in medium and high SNR regions where practical applications are operated, the average latency of the hybrid decoder is very similar to the BP decoder with *max iter*=60. More importantly, as shown in Fig. 5, the extra decoding gain (at least 0.2dB) provided by the hybrid decoder makes this 30% worst-case latency tolerable.

In addition, compared to a single BP decoder that has the same worst-case latency (*max iter*=315), the hybrid decoder has similar average decoding latency with extra 0.2dB decoding gain at the high SNR region (see Fig. 5). As a result, the hybrid decoder performs much better than a single BP decoder with large *max iter* when considering both latency and decoding performance. The similar advantage of hybrid decoder on this joint consideration also exists in the comparison with the SC decoder. From Fig. 5 and Fig. 6 it can be seen that the hybrid decoder has much shorter average decoding latency than SC decoder with 0.2dB gain at the medium SNR. In general, the proposed hybrid decoder achieves much shorter decoding latency with improved decoding performance.

## IV. Hardware Architecture and Analysis

Section III presented the hybrid decoder. A straightforward implementation of this decoder needs hardware resources for both SC and BP algorithms. However, after careful review of inherent decoding procedure of SC and BP algorithms, we propose a unified architecture that performs these two algorithms using the same hardware.

The unified architecture exploits the similarity of mathematicial forms of basic computations in SC and BP algorithms. As indicated in [9], the SC decoder consists of two basic computations that are described below:

$$\boldsymbol{f}(a,b)\approx\mathbf{sign}(a)\mathbf{sign}(b)\mathbf{min}(|a|,|b|) \quad (1)$$

$$\boldsymbol{g}(a,b)=a(-1)^{\hat{u}_{sum}}+b \quad (2)$$

Similarly, [12] showed that the BP decoder consists of the following two basic computations:

$$d=s*\mathbf{sign}(in_1)\mathbf{sign}(in_2+in_3)\mathbf{min}(|in_1|,|in_2+in_3|) \quad (3)$$

$$d=in_1+s*\mathbf{sign}(in_2)\mathbf{sign}(in_3)\mathbf{min}(|in_2|,|in_3|) \quad (4)$$

Comparing (1) and (3), it is found that (1) is the special case of (3) with $s$=1, $in_1=a$, $in_2=b$ and $in_3$=0. Similarly, (2) can be viewed as the special case of (4) when $s=\mathbf{sign}(b)$, $in_1=(-1)^{\hat{u}_{sum}}a$, $in_2=b$ and $in_3=b$. As a result, the unified basic computation units for hybrid decoder are developed in Fig. 7(a) and Fig. 7(b), respectively. From the figure it is seen that with proper selection of input signals, the unified computation blocks can perform the basic function of SC and BP decoders using the same hardware.

Fig. 7(c) and Fig. 7(d) show the example decoding scheme of single BP and SC decoders for $n$=8 polar codes, respectively. In order to perform the valid decoding scheme of the hybrid decoder, a configurable finite state machine (FSM) needs to be integrated to the control block of hybrid decoder to select the proper decoding scheme in different decoding phases.

Table I shows the estimated hardware performance for different polar codes decoders for (1024, 512) codes. Here the SC decoder is used as an 8-bit output version. It can be seen that the proposed hybrid decoder achieves 0.2dB gain than the single BP decoder with the same hardware performance. In addition, it has 10 times increase in throughput than the SC decoder with the same decoding performance at 4dB.

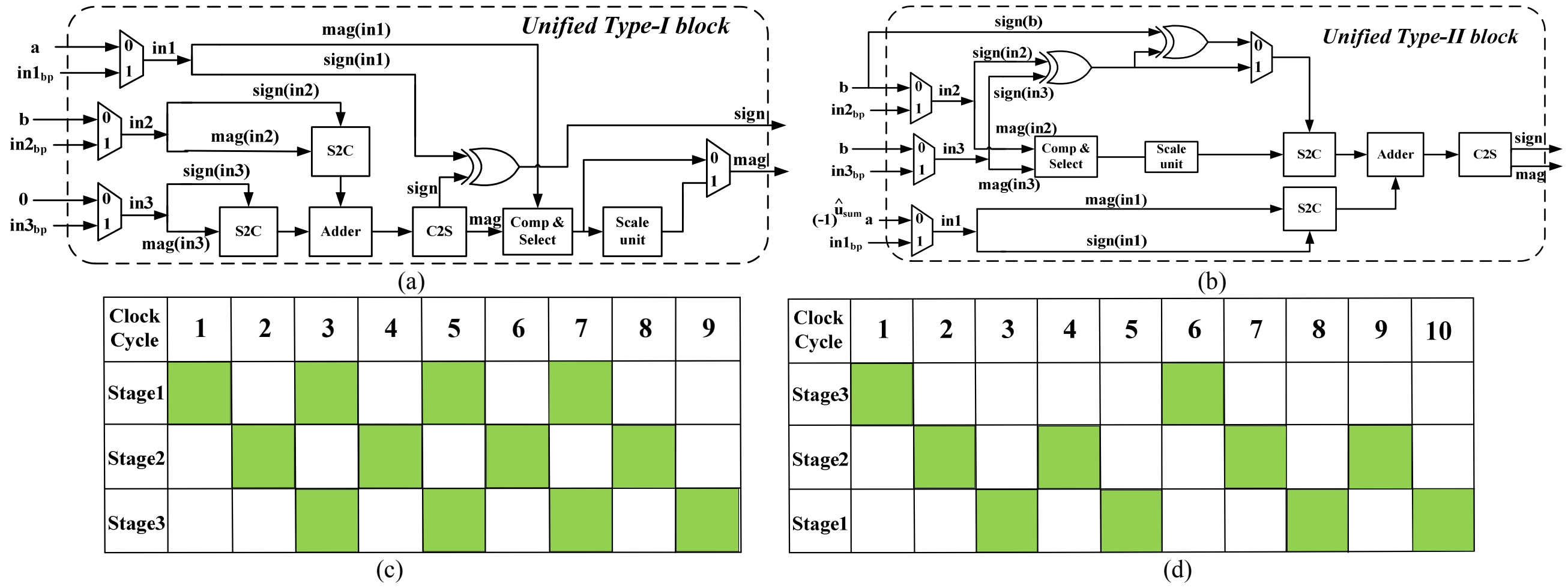

Fig. 7. (a) Unified Type-I block. (b) Unified Type-II block. (c) Decoding scheme of BP decoder. (d) Decoding scheme of SC decoder.

TABLE I. HARDWARE ESTIMATION OF DIFFERENT POLAR CODES DECODERS FOR (1024, 512) CODES

| Architecture | Hybrid (BP with *max iter*=60 + 8-bit output SC) | BP with *max iter*=60 | BP with *max iter*=315 | SC (8-bit output) |
|---|---|---|---|---|
| # of PE (normalized) | 5120 | 5120 | 5120 | 1024 |
| Critical path delay ($T_{adder}$) | 4 | 4 | 4 | 15(before retiming) 4(after retiming) |
| Average decoding latency@4dB (cycles) | 51 | 51* | 51 | 510† |
| Average decoding throughput@4dB (normalized) | 10 | 10 | 10 | 1 |
| Coding gain over BP with *max iter*=60 | >0.2dB | N/A | <0.05dB | >0.2dB |

*According to [15], the latency of BP decoder is approximately $2v+m$ cycles, where $v$ is the number of iterations, and $m$=log2$n$.

†According to [10], the latency of $2^K$-bit output SC/SCL decoder is $n/2^{K-2}$-2 cycles.

Although hybrid decoder has larger hardware complexity, the hardware efficiency, as the ratio between throughput and hardware, is larger than that of a single SC decoder. As a result, the proposed hybrid decoder achieves good error-correcting and hardware performances.

## V. CONCLUSION

This paper presents a hybrid decoding scheme for polar codes. With the use of concatenation of BP and SC decoders, the hybrid decoder can simultaneously achieve good error-correcting and hardware performance.